\documentclass[a4paper,12pt]{article}

\usepackage[english]{babel}
\usepackage[utf8]{inputenc}
\usepackage{amsmath}
\usepackage{graphicx}
\usepackage[colorinlistoftodos]{todonotes}
\usepackage[margin=1in]{geometry}
\usepackage[hidelinks]{hyperref}
\usepackage{amssymb}
\usepackage{stix}
\usepackage{gensymb}
\usepackage {tikz}
\usepackage{adjustbox}
\usepackage{multirow}
\usepackage{graphicx}
\usepackage{xcolor}
\usepackage{caption}
\usepackage{subcaption}
\usepackage{graphicx}

\usetikzlibrary {positioning}

\title{Attitudes and Latent Class Choice Models using Machine Learning}
\bigskip
\usepackage{float}
 \author{\textit{Lorena Torres Lahoz$^{1*}$}, \textit{Francisco Camara Pereira $^{1}$}, \\\textit{Georges Sfeir$^{1}$}, \textit{Ioanna Arkoudi$^{1}$},
\\\textit{Mayara Moraes Monteiro$^{1}$}, \textit{Carlos Lima Azevedo$^{1}$}
 \\
\textit{$^1$DTU Management, Technical University of Denmark}\\
\textit{Building 116, Bygningstorvet, 2800 Kongens Lyngby}\\
\textit{$^*$Corresponding author: ltola@dtu.dk}
}
\begin{document}
\maketitle
\section*{Abstract}
Latent Class Choice Models (LCCM) are extensions of discrete choice models (DCMs) that capture unobserved heterogeneity in the choice process by segmenting the population based on the assumption of preference similarities.
We present a method of efficiently incorporating attitudinal indicators in the specification of LCCM, by introducing Artificial Neural Networks (ANN) to formulate latent variables constructs. This formulation overcomes structural equations in its capability of exploring the relationship between the attitudinal indicators and the decision choice, given the Machine Learning (ML) flexibility and power in capturing unobserved and complex behavioural features, such as attitudes and beliefs. All of this while still maintaining the consistency of the theoretical assumptions presented in the Generalized Random Utility model and the interpretability of the estimated parameters. We test our proposed framework for estimating a Car-Sharing (CS) service subscription choice with stated preference data from Copenhagen, Denmark. The results show that our proposed approach provides a complete and realistic segmentation, which helps design better policies.
\subsection*{Keywords}
Machine learning, Latent Class Choice Models, Car-Sharing, Psychometric Indicators, Deep Learning.
\newpage
\section{Introduction}
Latent Class Choice Models (LCCMs) are extensions of discrete choice models (DCMs) that are typically implemented to capture unobserved heterogeneity in the choice process by segmenting the population based on the assumption of preference similarities. Traditionally, the latent classes are often defined by the socio-economic characteristics of the decision-makers \cite{Hess, Hess2, walker}.
 However, incorporating individuals' attitudes and/or perceptions, typically measured using psychometric indicators, into the specification of the latent classes can offer additional behavioural insights, allowing for a more realistic market segmentation that can help design more appropriate strategies and effective policies \cite{Motoaki, SubIndicator2, SubIndicator3}.

In this study, we explore a new method of efficiently incorporating attitudinal indicators in the specification of LCCM by relying on Machine Learning (ML) techniques. All this while preserving the benefits of the economic and behavioural interpretability of DCMs.

There are many examples in the literature of employing ML in DCMs, such as the ones reviewed in section 2.3. However, there is a lack of effective use of ML techniques for incorporating attitudinal variables into the model formulation. Even though such variables have been included in traditional DCMs, a complex interaction with the decision-making process should be expected \cite{latent}. We hypothesise that ML could be a good starting point to explore it, given its flexibility and power in capturing unobserved and complex interactions.
Therefore, we propose a model that includes latent classes and latent variables, the latter being defined as an Artificial Neural Network (ANN).

Finally, the developed model follows the generalized random utility model \cite{walker} theoretical structure, which provides us with a discrete representation of heterogeneity, and, therefore, higher-level behavioural insights into the composition of the latent classes. This has been shown to allow for more accurate market segmentation and policies that account for the different characteristics, preferences, and attitudes of individuals in each market segment (see for example \cite{Carlos1}, \cite{Carlos2}).

\section{Literature Review}
\subsection{Latent Class Choice Models (LCCM)}
Based on McFadden's model \cite{Nobel}, Walker and Ben-Akiva \cite{walker} presented a practical generalized random utility model with extensions for latent variables and classes. They extended basic Random Utility Model (RUM) to relax assumptions and enrich the model's capabilities. Latent classes refer to unobserved population groups, in which each individual has an associated probability of belonging to each group/class.\\ 
LCCM are a particular case of mixed logit models. Mixed logit models are defined as logit models where the coefficients of the utility function vary across decision-makers according to a given density $f(\beta)$ \cite{Mixed}. Therefore, unobserved heterogeneity is captured by assuming a parameter distribution over the population, where each individual has a slightly different behaviour toward the same attribute of the alternatives.\\
For the particular case of LCCM, the hypothesis is that there may be discrete segments of decision-makers that are not immediately identifiable from the data. Hence, the population can be probabilistically segmented into groups with different preferences and decision protocols or that behave differently towards a decision, signified by class-specific utility equations for each class. This segmentation could lead to more practical results, such as more efficient design policies \cite{walker}.\\
Although mixed logit models are highly flexible and can approximate any random utility model \cite{McFadden}, they are more challenging to interpret and use in practice as they do not identify the causes for the variations of taste among decision-makers. On the other hand, evidence in the literature suggests that latent class models are very convenient, flexible, and intuitive to account for taste heterogeneity in discrete choice models  \cite{hurtubia}.\\
A critical part of LCCMs is specifying the class membership model since it is not defined beforehand. Traditionally, straightforward logit formulations are employed since they allow for a closed-form expression of the probability to belong to each class. Moreover, they typically base their segmentation on only socio-demographic variables \cite{Hess, Hess2, walker}.\\
In the following sections, we will explore different class membership formulations that include information about attitudinal variables. As we have mentioned, incorporating individuals' attitudes and/or perceptions into the specification of the latent classes helps construct more realistic population segments, allowing for more accurate and successful policies \cite{hurtubia, Motoaki}.

\subsection{Latent Psychosocial Constructs in DCM}
Apart from the attributes of the alternatives and the characteristics of the decision-maker, according to psychological theory, there exist more complex, unobserved constructs that may have a relevant effect on how we make choices. Some of these latent factors can be the decision maker's lifestyle, attitudes, or perceptions \cite{McFadden2}.\\
Psychometric indicators measure the effect of unobserved attributes on individuals' preferences on topics related to the choice. They are additional information that can helps specify and estimate latent classes.
Atasoy et al. \cite{Indicators1} estimated an LCCM where psychometric indicators are included in the maximum likelihood estimation to improve the model's accurancy. The psychometric indicators were modelled as conditional on the latent class $k$. Therefore, the item-response probability of observing indicator $I_n$ was given by $P_n(I_n|k)$ that was defined as a parameter jointly estimated with the choice and the class membership model. The model evidence that includes the psychometric indicators allows for richer analysis and generates significantly different estimates for the class membership model.\\ 
Another approach for including attitudinal variables in DCM models can be found in work developed by Hurtubia et al. \cite{hurtubia}. Psychometric indicators are introduced by assuming that the probability of giving an agreement level to an attitudinal question also depends on the respondent's class. In this case, they use an ordinal logit approach to model the response probability  $P_n(I_k|k)$, since the responses to the indicators consist of a few ordered integer values corresponding with the level of agreement with the statement on a Likert scale \cite{Likert}. The advantage of this formulation is the close form of the ordinal logit used to measure the indicators, which allows for a more straightforward estimation procedure, where the choice and the response to the indicators are estimated together.\\
Another study \cite{Indicators3} has gone further by including latent variables and latent classes, to model the relationship between normative beliefs, modality styles, and travel behaviour.
Normative beliefs refer to the individual's perception of the opinion of others concerning a specific behaviour, while modality style describes the part of an individual's lifestyle characterized by using a particular set of travel modes. The study shows evidence of associations between travel behaviour and different latent psychosocial constructs.
Their model exemplifies how the socio-psychological approach, which originated in social psychology, can extend conventional travel behaviour analysis. However, due to the model's complexity, sequencial estimation procedure is employed, encountering identification issues for some coefficients.\\
Finally, Alonso-González et al.\cite{Indicators4} presented an exploratory factor analysis and latent class cluster analysis of attitudinal variables performed to divide individuals into groups concerning their inclination to adopt MaaS (Mobility as a Service). Even though the model is enriched with a series of covariates referring to socioeconomic, mobility, and technology-related characteristics, they do not improve the model; they only help cluster identification. Attitudinal variables are the ones defining the structure of the latent classes.\\
The previous examples have highlighted the importance and relevance of including attitudes, preferences, and beliefs in DCMs, especially in LCCM. They have proved to be relevant for improving the model estimation and creating more realistic and meaningful classes.
\subsection{Machine Learning for Taste Heterogeneity in DCM}
In recent years, the use of ML techniques has increased, mainly due to their power to improve prediction accuracy. However, one of the main criticisms of ML techniques vs. econometric models is that it provides results that are less interpretable. Thus, in the transportation field, researchers have also focused on providing interpretable and meaningful estimates from ML applications, so that they can be more useful for travel analysis and policy decisions.\\
 Some interesting works are, for example, Sifringer et al. \cite{Embeddings2}, which, based on previous works \cite{Previous1},\cite{Previous2},\cite{Previous3}, included an additional term in the utility layer called the Learning Term, which was estimated separately with a Dense Neural Network, using a priori disregard explanatory variables. This model outperformed previous formulations while keeping the interpretability of the other terms of the utility. Following this idea and based on Pereira \cite{Pereira} previous work, Arkoudi et al \cite{Ioanna} proposed an embedding encoding for the socio-characteristic variables that provided a latent representation of these variables in concordance with individuals' choices. Moreover, they included a Learning Term that increased the model's accuracy. However, this embedding formulation does not consider information about attitudinal indicators since these variables cannot be directly included in the utility specifications. Furthermore, none of these works provided latent classes or variables formulations.\\
 The work developed by Han et al \cite{ML1}, proposed a nonlinear LCCM by using a neural network to specify the class membership model. In their application, a model with eight latent classes outperformed the previous ones in prediction accuracy. Due to non-linearity, the proposed LCCM was less transparent and lost interpretability.
The work of Sfeir et al. presented two model formulations for the construct of latent class choice models using Gaussian process \cite{Georges2} and Mixture models \cite{George}, while Wong et al. \cite{ML2} employed a generative non-parametric ML approach, a Boltzmann machine, to estimate latent variables. Although these works allowed for more flexibility in the definition of these latent constructs, none of them relied on attitudinal indicators' information.\\
To summarize, the theory of LCCM has been widely explored and successfully employed for many decades, and more recently, ML techniques have provided practical solutions for situations where the traditional discrete choice models were more constrained, improving their estimation and flexibility. However, the employment of ML for incorporating attitudinal information in DCM is still in the earliest steps of research, given its theoretical and practical underlying complexity.

\section{Methodology}
\begin{figure}[H]
    \centering
    \includegraphics[width = 1
    \textwidth]{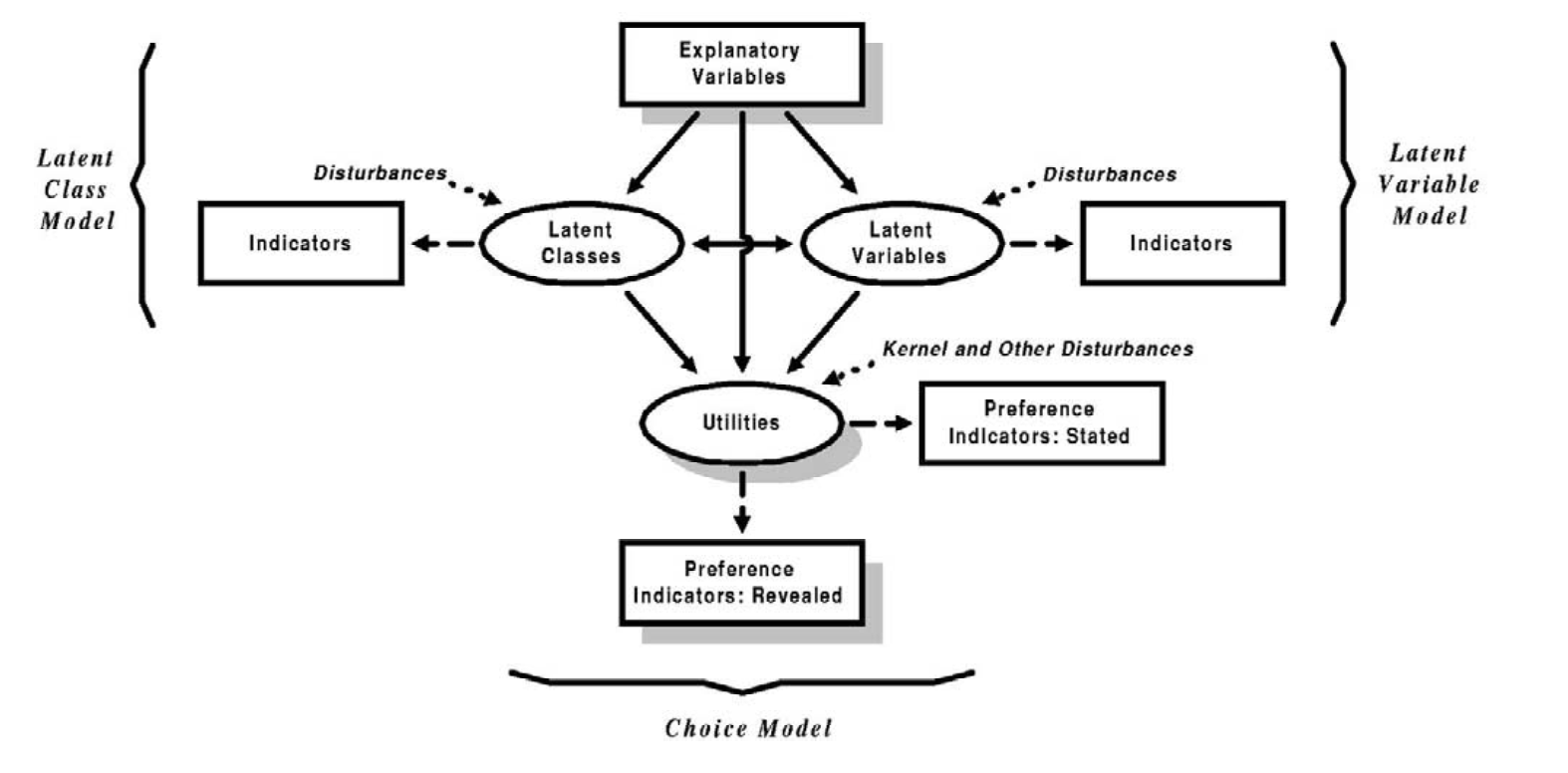}
    \caption{Generalized random utility model \cite{walker}}
    \label{M0}
\end{figure}

In this section, we present our model formulation following the generalized random utility model structure (Figure \ref{M0}) presented by  Walker et al., 2002 \cite{walker} and the added modifications for the incorporation of latent variables through the employment of ANN in the latent class model. \\
In Figure 1, solid arrows represent structural equations (cause-effect relationships), and dashed arrows represent measurement equations (relationship between observable indicators and the underlying latent variable).

\subsection{Latent Class Choice Models (LCCM)}
LCCMs are composed of two sub-models: \textit{a class membership model} and a \textit{class-specific choice model}. The former computes the probability of an individual $n$ belonging to a certain class, while the latter assigns the probability of choosing each alternative, given that individual $n$ belongs to a certain class $k$.\\
Typically, in  the \textit{class membership model} is define as logit, where the utility of an individual $n$ belonging to class $k$ can be expressed as follows:
\begin{equation}
    U_{nk}= ASC_k + Q_n\gamma_k + \upsilon_{nk}
\label{1}
\end{equation}
where $ASC_K$ is the alternative-specific value for each class $k$, $Q_n$ is a vector containing the characteristics of individual \textit{n}, $\gamma_k$ is a vector of the unknown parameters that need to be estimated and $\upsilon_{nk}$ is an error term that is assumed to be independently and identically distributed Extreme Value Type I over individuals and classes. Thus, the probability that individual $n$ belongs to class $k$ given their characteristics is the logit formula:
\begin{equation}
   P(q_{nk}| Q_n,\gamma_k)= \frac{e^{Q_n\gamma_k}}{\sum^K_{k'=1}e^{Q_n\gamma_{k'}}}
\end{equation}
where $q_{nk}= 1$ if the individual belongs to class $k$ and 0 otherwise.\\
As for the \textit{class-specific choice model}, it returns the probability that an individual $n$ chooses a specific alternative, given that he/she belongs to a certain class $k$.\\
The utility of individual \textit{n} choosing alternative \textit{i}, given that he/she belongs to class \textit{k} is formulated as:
\begin{equation}
    U_{ni|k}= V_{ni|k} + \epsilon_{ni|k}= X_{ni}\beta_k + \epsilon_{ni|k}
\end{equation}
where $V_{nit|k}$ is the observed part of the utility, and $\epsilon_{nit|k}$ is an independent and normally distributed Extreme Value Type I error term distributed over individuals, alternatives and classes. $X_{ni}$ is the vector of observed attributes of alternative \textit{i}  and normally includes an alternative-specific constant, and $\beta_k$ is the vector of unknown parameters that need to be estimated.\\
Therefore, the probability that individual $n$ chooses alternative $i$, conditioned on class $k$, is formulated as:
\begin{equation}
   P(y_{ni}| X_{ni}, q_{nk}, \beta_{k})= \frac{e^{V_{ni|k}}}{\sum^J_{j'=1}e^{V_{nj'|k}}}
\end{equation}
where $y_{ni}$ is 1 if individual $n$ chooses alternative i and 0 otherwise, and $J$ is the total number of alternatives. \\
Conditional on the class, the probability of observing $y_{n}$ is formulated as:\\
\begin{equation}
   P(y_{n}| X_{n}, q_{nk}, \beta_{k})= \prod^{J}_{j=1}(P(y_{nj}|X_{nj}, q_{nk}, \beta_k))^{y_{nj}}
\label{5}
\end{equation}
where $y_n$ is a $J$ vector of all choices $y_{nj}$ of an individual $n$ and $X_n$ is a $J$ vector of $X_{nj}$.\\
The unconditional probability of observing $y_n$ can be calculated by mixing the previous conditional choice probability with the probability of belonging to each class $k$, as follows: 
\begin{equation}
   P(y_{n})= \sum^{K}_{k=1}P(q_{nk}|Q_{n}, \gamma_k)P(y_{n}| X_{n}, q_{nk}, \beta_{k})
\end{equation}
Finally, the likelihood over all individuals $N$, considering independence between them, is:
\begin{equation}
   P(y)= \prod^{N}_{n=1}\sum^{K}_{k=1}P(q_{nk}|Q_{n}, \gamma_k)P(y_{n}| X_{n}, q_{nk}, \beta_{k})
\end{equation}
In section \ref{base}, we employ this LCCM formulation as a baseline for our new model.
\subsection{Latent Variables and Latent Classes with ANN for Incorporation of Attitudinal Indicators}
\begin{figure}[H]
    \centering
    \includegraphics[width = 0.9
    \textwidth]{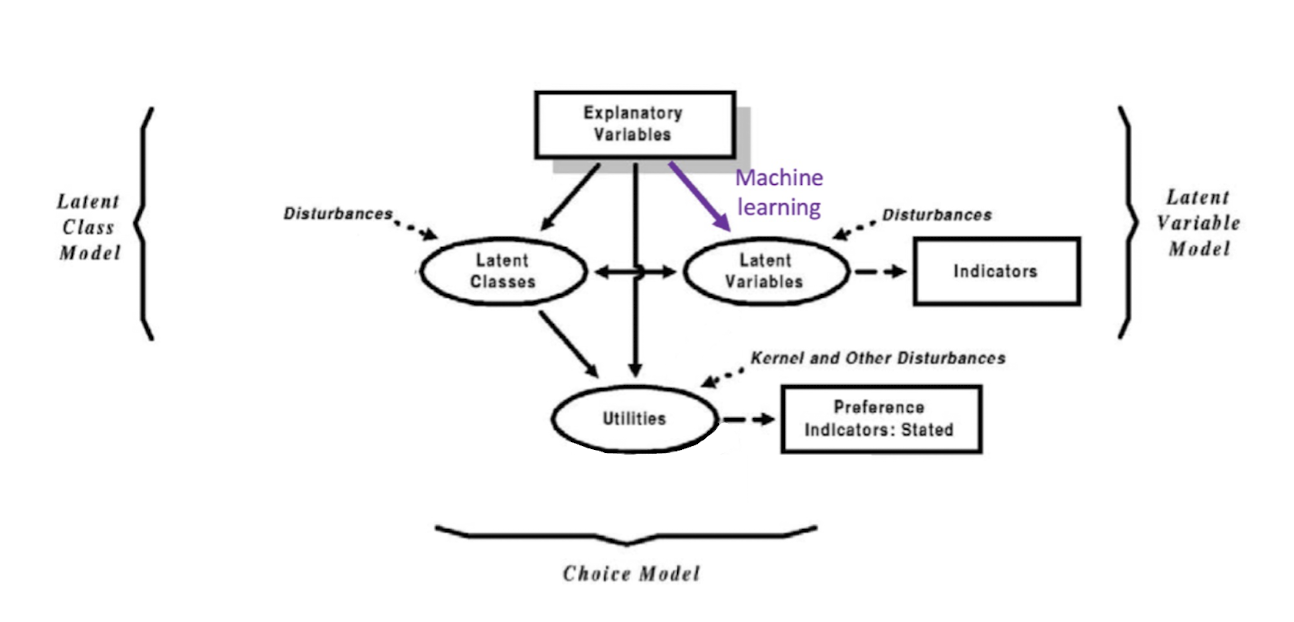}
    \caption{Graphical representation of the model formulation}
    \label{M1}
\end{figure}

The straightforward solution for incorporating the attitudinal indicators in DCMs would include their responses directly into the utility of the latent classes. However, there are three main reasons to avoid this approach.
Firstly, this type of data is not always available for forecasting since these types of surveys are usually expensive and time-consuming. Therefore, the model can predict the class membership model probability even when no information about attitudinal indicators is provided. The second reason is that the multicollinearity inherent in the responses makes it difficult to estimate the unknown parameters correctly and can lead to inconsistent estimates. In this sense, even a low-dimensional representation of the indicators, e.g., factor analysis, cannot be included directly in the utility unless the correlated error terms are added.
Lastly, and more importantly, there is no apparent justification that such attitudinal information can uncover some causal relationship directly with behaviour. Instead, the literature points out that these indicators seem to be the consequences of some latent constructs, which combined can represent individuals' attitudes \cite{MIT}.\\
We follow the generalized random utility model structure presented by \cite{walker} to address the previously mentioned issues. Moreover, we include the information on the attitudinal indicators by employing an ANN to formulate the latent variables constructs, which provides more flexibility in how the latent variables are formulated. The graphical representation of the proposed formulation is shown in Figure \ref{M1}.\\
Finally, we decided not to include the arrow between the latent variables and the utilities, and the revealed preference indicators, since we did not find it relevant to our concrete application. However, these relations can be included for other cases by following the same principle as in Figure \ref{M0} \cite{walker}.

\subsubsection{Latent classes with latent variables formulation}
The utility of the class membership model can be written as:
\begin{equation}
\begin{split}
U_{nk}= V_{nk} + \upsilon_{nk}
\end{split}
\label{8}
\end{equation}
where $V_{nk}$ is the representative utility of individual $n$ belonging to class $k$ and, as defined in the traditional LCCM, $\upsilon_{nk}$ is the error term that is assumed to be independent and identically distributed Extreme Value Type I over individuals and classes ans defined in \ref{1}.\\
In this case, we define $V_{nk}$ as:
\begin{equation}
\begin{split}
V_{nk} = ASC_k + Q_n\gamma_k + r_{n}\delta_{k} + \omega_nb_k
\end{split}
\label{9}
\end{equation}
where $ASC_k$ is the alternative-specific value for class $k$, $Q_n$ is the vector containing socio-characteristics of individual \textit{n}, and $\gamma_k$, the vector of unknown parameters that need to be estimated for each class $k$. In addition, $r_{n}$ is the vector of latent variables for individual $n$ and $\delta_{k}$ the corresponding vector of unknown parameters specific to class $k$, that also need to be estimated. Finally, $\omega_n$ is an individual-specific constant  with its corresponding coefficient $b_k$ for each class $k$.\\
The term $\omega_n$ represents the individual variation of all the latent variables caused by the variance of their underlying distributions. It is formulated as a one-layer ANN that gets activated by the ID of each decision maker in the train set ($Id_n$ is 1 for decision-maker $n$ and 0 otherwise)
\begin{equation}
   \omega_n = \sum_1^Nw_{1n}^{(1)}Id_n
\end{equation}
where $w_{1n}^{(1)}$ are the weights of the layer. Since only the difference between classes matters, we introduce the corresponding $b_k$ term for each class, which accounts for the effect of the latent variables variation for each of the latent classes.\\
Given the distribution of the error term ($\upsilon_{nk}$), the probability $P(q_{nk}|Q_{n}, \gamma_k, r_n, b_{k})$ can be expressed as:
\begin{equation}
   P(q_{nk}|Q_{n}, \gamma_k, r_n, b_{k})= \frac{e^{V_{nk}}}{\sum^K_{k'=1}e^{V_{nk'}}}
\label{11}
\end{equation}
As it is standard practice in class membership models, the utility of one class is set to zero due to the parameters' identification since only the differences in utility matter.\\

\subsubsection{Latent variables formulation as ANN}
The novelty of this work is the employment of ANN for the construction of latent variables. We propose a non-linear relationship between the socio-characteristics of the individuals and the latent constructs by employing two densely connected layers. It can be expressed as follows:
\begin{equation}
r_{zn} =a_2(\sum_{h=0}^H w_{zh}^{(2)}a_1(\sum_{m=0}^Mw_{hm}^{(1)}Q_{mn}))
\end{equation}
where $M$ is the number of socio-demographic variables used to predict the answer to the indicators, and $H$ is the number of hidden units in the hidden layer.
$w_{hm}^{(1)}$ are the weights of the first layer, and $a_1$ represents the first activation function defined as a Rectified Linear Unit (ReLU) ($a_1(x)=max(0,x)$); for the second layer, a linear activation function is applied $a_2(x) = x$, and the weights are represented by $w_{zh}^{(2)}$. By adding an extra input $Q_{0n}$,  which is set to one and extending the sum to go from zero, we avoid writing the intercept term.\\
The number of latent variables, defined as $Z$, the number of hidden neurons in the hidden layer, $H$, and the number of densely connected layers are dimensions that the researcher can tune since they are not observed in the data. Employing different values for these hyperparameters for a different model application does not invalidate the proposed formulation and can help achieve higher model prediction for the test data.\\
The formulation presented is based on the hypothesis that the socio-characteristics of the individuals define the latent variables. Moreover, these latent constructs influence the response to specific attitudinal indicators.

\subsubsection{The ordinal measurement model}
\begin{figure}[H]
    \centering
    \begin{subfigure}[b]{0.6\textwidth}
    \includegraphics[angle=90,origin=c, width = 1 \textwidth
   ]{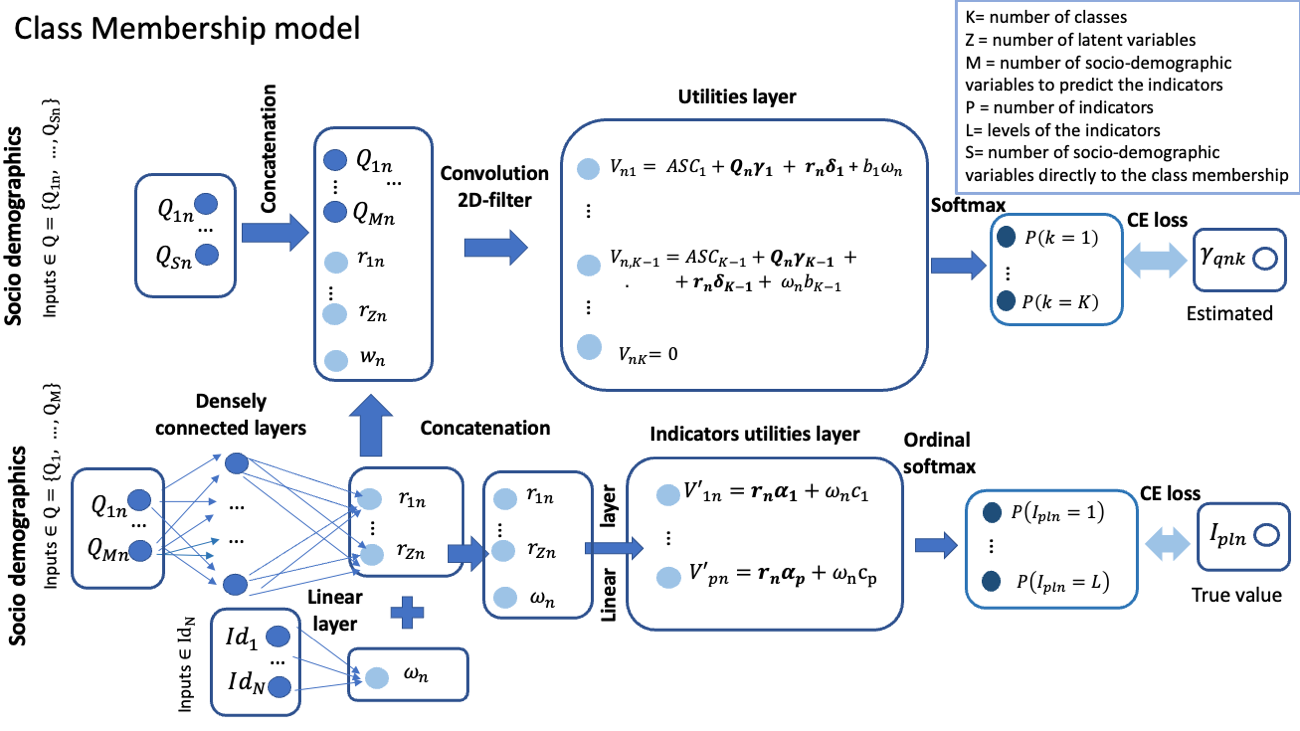}
    \end{subfigure}
    \begin{subfigure}[b]{0.3\textwidth}
    \includegraphics[angle=90,origin=c, width = 1 \textwidth
   ]{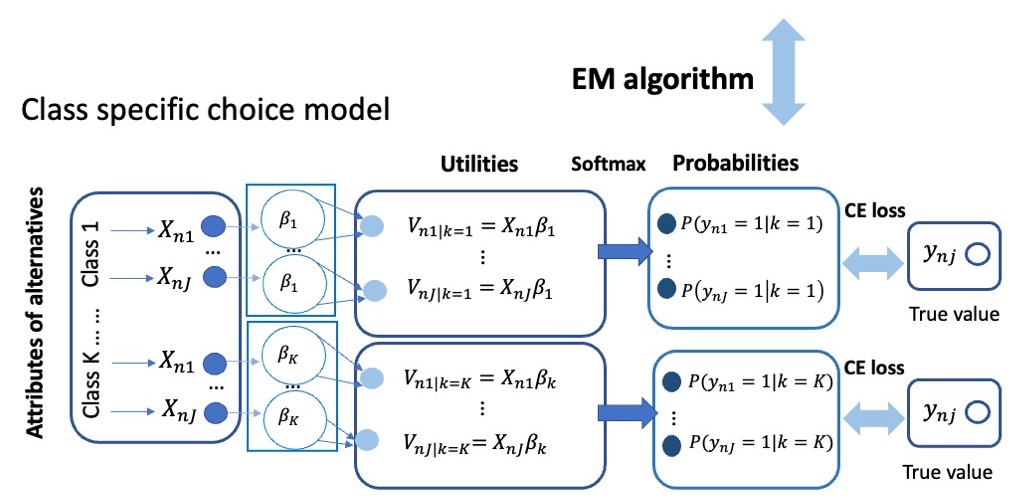}
    \end{subfigure}
    \caption{Model architecture\protect\footnotemark}
    \label{M3}
\end{figure}

We focus on the case where indicators take the form of questions that receive an ordered response, as in Likert scales \cite{Likert}. Each option of the scale represents a level of agreement with a particular statement. Thus, we define the utility of individual $n$ for indicator $p$, as a measurement of the level of agreement with its statement, and we formulated it as:
\begin{equation}
   U_{pn}= V_{pn} + \nu_{pn} = r_{n}\alpha_p + c_p\omega_n +\nu_{pn}
\end{equation}
where $V_{pn}$ is the representative utility of individual $n$ to indicator $p$ and $\nu_{pn}$ is the error term that is assumed to be independently and identically distributed Extreme Value Type I over individuals and indicators. $r_{n}$ is a vector of length $Z$ containing the latent variables of individual $n$, $\alpha_p$ is also a vector of size $Z$ with its corresponding parameters to be estimated. $\omega_n$ is the individual-specific parameter estimated together with the latent class model, and $c_p$ is its corresponding coefficient for each indicator $p$ .\\
Therefore, the probability that individual $n$ answers with a certain level of agreement $l$ to indicator $p$ is expressed as:
\begin{equation}
   P(I_{pln}=1|r_n, \alpha_p, c_p, \omega_n)= P(\tau^p_{l-1} < U_{pn} < \tau^p_l)
\label{14}
\end{equation}
where we define $I_{pln}$ as 1 if individual $n$ answers with a level of agreement $l$ to indicator $p$ and 0 otherwise. $\tau^p_l$ are strictly increasing class-specific thresholds that define an ordinal relation between the utility $U_{pn}$ and the level of agreement to indicator $p$.\\
Finally, the probability of individual $n$ providing an answer $l$ to indicator $p$ can be computed as an ordinal softmax:
\begin{equation}
\begin{split}
   P(I_{pln}=1|r_n, \alpha_p, c_p, \omega_n)=  P( \tau^p_{l-1} < U_{pn} < \tau^p_l) = & P( \tau^p_{l-1} < V_{pn}+ \nu_{vp} < \tau^p_l) = \\  = Prob(\nu_{vp} < \tau^p_l -V_{pn}) - P( \nu_{vp} < \tau^p_{l-1} - V_{pn})=
   &= \frac{e^{\tau^p_l -V_{pn}}}{1+ e^{\tau^p_l -V_{pn}}} - \frac{e^{\tau^p_{l-1} - V_{pn}}}{1+ e^{\tau^p_{l-1} - V_{pn}}}
   \end{split}
\label{15}
\end{equation}
where one threshold per indicator is set to zero, as only the difference between them matters.\\
\\
The final model formulation is presented in Figure \ref{M3}.
Maximizing the log-likelihood with respect to the unknown parameters will not lead to a closed-form solution. Thus, the estimation becomes more complex and challenging as empirical singularity issues could appear at some iterations \cite{Train}. To overcome these problems, we employ the Expectation-Maximization (EM) algorithm (Dempster et al., 1977 \cite{EM}), a powerful method used for maximizing the likelihood in models with latent variables.
\footnotetext{The expressions without simplifying for $P(k=1)$ and $P(k=K)$ can be found in equation (\ref{11}), for $P(I_{pnl}=1)$ and $P(I_{pnl}=L)$ in equation (\ref{15}) and for $P(y_{nj}=1|k=K)$ and similars in equation (\ref{5}).}
\subsubsection{EM algorithm}
We estimate all components of the proposed model together since simultaneous estimation usually leads to more efficient estimates than sequential estimation \cite{Sequencial2}, \cite{McFadden2}.\\
For this purpose, we employ the EM algorithm, which combines an expectation step with a maximization one until convergence is reached. We start by randomly initializing the unknown parameters. After that, we estimate the expected values of the latent variables (E-step) using Bayes' theorem. Then, we update the unknown parameters' values using log-likelihood maximisation (M-step). Finally, we evaluate the log-likelihood with the updated values of the unknown parameters and check for convergence. We return to the E-step until convergence is reached. In this case, the convergence is defined by the number of iterations.\\
We can rewrite the joint likelihood assuming that the clusters are observed as follows:
\begin{equation}
\begin{split}
 P(y, q, Q, X)= &\prod^N_{n=1}\prod_{k=1}^K
\left[\frac{e^{V_{nk}}}{\sum^K_{k'=1}e^{V_{nk'}}}\right]^{q_{nk}} \\ & X\prod^{N}_{n=1}\prod^{K}_{k=1}\prod^{J}_{j=1}\left[\frac{e^{V_{nj|k}}}{\sum^J_{j'=1}e^{V_{nj'|k}}} \right]^{y_{nj}q_{nk}}
\end{split}
\end{equation}
The function is divided into two parts when we take the logarithm of the likelihood:
\begin{equation}
\begin{split}
LL= \sum^N_{n=1}\sum_{k=1}^K q_{nk}log
\left[\frac{e^{V_{nk}}}{\sum^K_{k'=1}e^{V_{nk'}}}\right] +& \sum^{N}_{n=1}\sum^{K}_{k=1}\sum^{J}_{j=1}y_{nj}q_{nk}log\left[\frac{e^{V_{nt|k}}}{\sum^J_{j'=1}e^{V_{n't|k}}} \right]
\end{split}
\label{18}
\end{equation}
this equation can be solved by equating its derivatives to zero with respect to each of the unknown parameters of the choice model if $q_{nk}$ is assumed to be known.\\
In order to find the initial values of $q_{nk}$, we estimate the expectation of $q_{nk}$ (E-step) using Bayes's theorem:
\begin{equation}
\begin{split}
E[q_{nk}] = \gamma_{qnk} = \frac{\frac{e^{V_k}}{\sum^K_{k'=1}e^{V_{k'}}}\prod^{J}_{j=1}\left[\frac{e^{V_{nj|k}}}{\sum^J_{j'=1}e^{V_{nj'|k}}}\right] ^{y_{nj}}}{\sum_{k'=1}^K\left[\frac{e^{V_k}}{\sum^K_{k'=1}e^{V_{k'}}}\prod^{J}_{j=1}\left[\frac{e^{V_{nj|k}}}{\sum^J_{j'=1}e^{V_{nj'|k}}} \right]^{y_{nj}}\right]}
\end{split}
\label{20}
\end{equation}
where $\gamma_{qnk}$ is considered as the posterior probability of the classes.\\
Combining equations \eqref{18} and \eqref{20}, we can formulate the expected value of the log-likelihood as:
\begin{equation}
E(LL)=\sum^{N}_{n=1}\sum^{K}_{k=1}\sum^{J}_{j=1}y_{nj}\gamma_{qnk}log\left[\frac{e^{V_{nj|k}}}{\sum^J_{j'=1}e^{V_{nj'|k}}} \right] + +\sum^N_{n=1}\sum_{k=1}^K \gamma_{qnk}log
\left[\frac{e^{V_k}}{\sum^K_{k'=1}e^{V_{k'}}}\right]
\end{equation}
By setting the derivatives of the expected log-likelihood to zero with respect to the unknown parameters $\beta_{k}$ of the choice model, we can estimate them with the following equation:
\begin{equation}
\beta_{k}= argmax_{\beta_k}\sum_{n=1}^N\sum_{j=1}^J y_{nj}\gamma_{q_{nk}}log\left [\frac{e^{V_{nj|k}}}{\sum_{j'=1}^Je^{V_{nj'|k}}}\right]
\label{beta1}
\end{equation}
where no closed-form solution can be obtained. Thus, we employ the gradient-based numerical optimization method BFGS \cite{Nocedal}.\\
For the unknown parameters of the class membership, back-propagation with cross-entropy loss is applied since the parameters are encoded inside a neural network. As presented in \cite{Embeddings2}, minimizing the cross entropy loss is equivalent to maximizing the log-likelihood function, which allows us to compute the Hessian matrix and, therefore, provides useful post-estimation indicators such as the standard deviation or the confidence intervals of the parameters of the model.\\
After convergence is reached, we can calculate the probability of observing a vector of choices $y$ for all individuals as follows:
\begin{equation}
    P(y)= \prod_{n=1}^N\sum_{k=1}^K P(q_{nk}|Q_{n}, \gamma_k, r_n, b_{k})\prod_{j=1}^J\left( P(y_{nj}|X_{nj}, q_{nk}, \beta_k) \right)^{y_{nj}}
\end{equation}
To benchmark our proposed model we use the formulation presented in \cite{walker} as baseline. One can represent it in graphical representation, similarly as in \ref{M3}, with the difference that the relation between the indicators' utility and the decision-makers' socio-demographic characteristics is linear.

\section{Data}  
The data used in this study comes from a survey carried out for the Share-More project \footnote{https://eitum-sharemore.net.technion.ac.il/}, which aimed to analyse individual preferences towards car-sharing (CS) services under different personalized incentive schemes.\\
The data was collected through a tailor-made online survey (available in both web and mobile versions), from July $16^{th}$ to August $6^{th}$ 2020, in Copenhagen (CPH). Respondents were recruited through panels, and the eligibility criteria were being more than 18 years old and having a valid driver's license. The survey yielded a total of  542 complete answers and was available in English and Danish. The average duration was around 15 minutes. It included five relevant parts:
\begin{enumerate}
    \item A brief introduction to the project and its objective. 
    \item A survey on the subject's socio-demographic characteristics ($Q_n$), including experience with CS. 
    \item A survey addressing questions regarding the respondents' attitudes toward private and CS. The respondents’ perceptions in regard to car ownership and CS usage were measured by their level of agreement with relevant statements using a 5-point Likert scale. The statements employed are presented in Table \ref{Ind}.
    \item A survey with questions related to preferences about CS incentives.
    \item A Stated Preference (SP) experiment with different options for CS plans, taking into account possible incentive schemes for CPH.
\end{enumerate}
\begin{table}[H]
\centering
\begin{tabular}{|c|l|}
\hline
\multicolumn{1}{|l|}{\textbf{Indicator}} & \textbf{Description} \\ \hline
$I_{1}$ & I feel stressed when driving \\ \hline
$I_{2}$ & It is difficult to find parking \\ \hline
$I_{3}$ & Owning and using a car is a big expense \\ \hline
$I_{4}$ & It’s easy for me to conduct my daily trips without a private car \\ \hline
$I_{5}$ & Driving a car is the most convenient way to move around \\ \hline
$I_{6}$ & For me the car is a status symbol \\ \hline
$I_{7}$ & I am worried about the environmental footprints of my car \\ \hline
$I_{8}$ & Car sharing is more affordable than owning my own car \\ \hline
$I_{9}$ & Car sharing is a more environmentally friendly alternative to car ownership \\ \hline
$I_{10}$ & By using car-sharing, I do not have to deal with vehicle maintenance and repair \\ \hline
$I_{11}$ & \begin{tabular}[c]{@{}l@{}}Thanks to car sharing, I can save from the fuel, taxes, insurance and parking \\ expenses associated with private vehicle ownership\end{tabular} \\ \hline
$I_{12}$ & Having access to different types of vehicles is an important advantage \\ \hline
$I_{13}$ & Using car-sharing service can lower my transport expenses \\ \hline
$I_{14}$ & Using car-sharing whenever I need it can make my life easier \\ \hline
$I_{15}$ & I would not need to buy a car because I have car-sharing \\ \hline
$I_{16}$ & Car-sharing is more convenient than public transport \\ \hline
$I_{17}$ & I wouldn’t mind sharing my personal car with other people \\ \hline
\end{tabular}
\caption{Attitudinal indicators employed in the survey}
\label{Ind}
\end{table}

\begin{figure}[H]
    \centering
    \includegraphics[width = 0.9
    \textwidth]{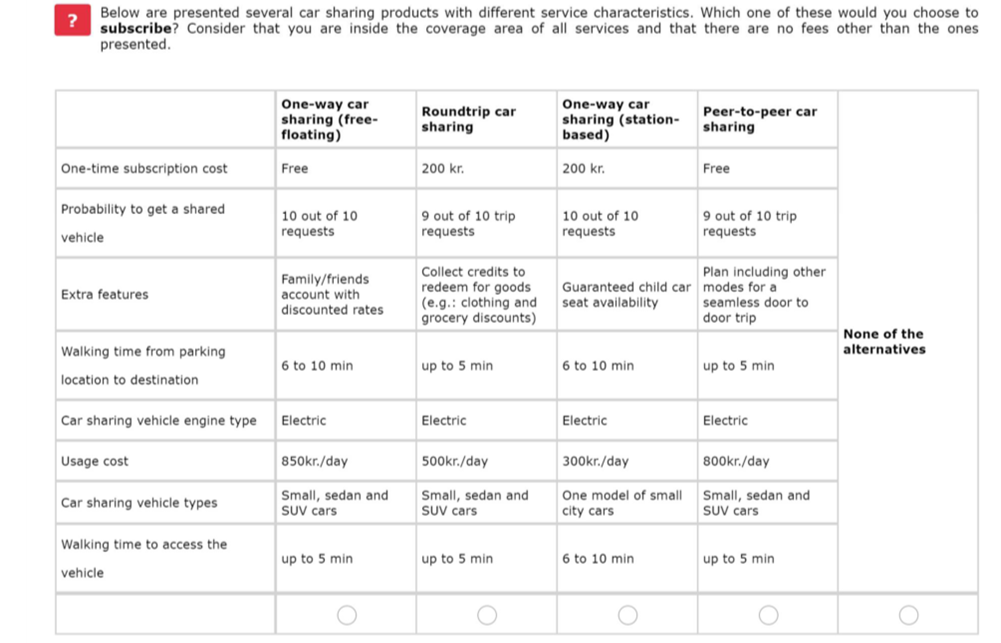}
    \caption{Example of choice task presented to respondents}
    \label{Example}
\end{figure}

 The sample socio-characteristics ($Q_n$) are described in Table \ref{tab:Sample}. More than 90\% of respondents stated to be aware of CS services, which indicates that they are well-known. The sample is quite balanced regarding gender and proportionally representative of the population in terms of age. Most of the respondents live in the city centre and are employed. For the level of education, more than 60\% of the population have at least a bachelor's degree. Moreover, most respondents live in households of 1 or 2 members, up to one car and have two or more bikes at home. Given the income level, most respondents earn around average (350.000 kr./year) or above, which is likely related to the high level of education of the sample.
 \begin{table}[H]
\centering
\begin{tabular}{|cl|c|c|}
\hline
\multicolumn{2}{|c|}{} & \textbf{Total} & \textbf{\%} \\ \hline
\multicolumn{1}{|c|}{\multirow{3}{*}{\textbf{Car -sharing membership status}}} & Car-sharing member & 96 & 17.68 \\ \cline{2-4} 
\multicolumn{1}{|c|}{} & Past-car sharing member & 64 & 11.79 \\ \cline{2-4} 
\multicolumn{1}{|c|}{} & Non-car-sharing member & 383 & 70.53 \\ \hline
\multicolumn{1}{|c|}{\multirow{2}{*}{\textbf{Car-sharing awareness}}} & Yes & 490 & 90.24 \\ \cline{2-4} 
\multicolumn{1}{|c|}{} & No & 53 & 9.76 \\ \hline
\multicolumn{1}{|c|}{\multirow{3}{*}{\textbf{Gender}}} & Man & 267 & 49.17 \\ \cline{2-4} 
\multicolumn{1}{|c|}{} & Woman & 275 & 50.64 \\ \cline{2-4} 
\multicolumn{1}{|c|}{} & Prefer not to say & 1 & 0.18 \\ \hline
\multicolumn{1}{|c|}{\multirow{5}{*}{\textbf{Age}}} & 18-30 & 146 & 26.89 \\ \cline{2-4} 
\multicolumn{1}{|c|}{} & 31-40 & 88 & 16.21 \\ \cline{2-4} 
\multicolumn{1}{|c|}{} & 41-50 & 97 & 17.86 \\ \cline{2-4} 
\multicolumn{1}{|c|}{} & 51-60 & 88 & 16.21 \\ \cline{2-4} 
\multicolumn{1}{|c|}{} & More than 60 & 124 & 22.84 \\ \hline
\multicolumn{1}{|c|}{\multirow{4}{*}{\textbf{Place of residence}}} & City center & 235 & 43.28 \\ \cline{2-4} 
\multicolumn{1}{|c|}{} & Suburbs & 190 & 34.99 \\ \cline{2-4} 
\multicolumn{1}{|c|}{} & Another city in the metropolitan region & 71 & 13.08 \\ \cline{2-4} 
\multicolumn{1}{|c|}{} & Outside the metropolitan region & 47 & 8.66 \\ \hline
\multicolumn{1}{|c|}{\multirow{6}{*}{\textbf{Employment status}}} & Student & 74 & 13.63 \\ \cline{2-4} 
\multicolumn{1}{|c|}{} & Employed & 354 & 65.19 \\ \cline{2-4} 
\multicolumn{1}{|c|}{} & Unemployed & 12 & 2.21 \\ \cline{2-4} 
\multicolumn{1}{|c|}{} & On leave & 7 & 1.29 \\ \cline{2-4} 
\multicolumn{1}{|c|}{} & Retired & 100 & 18.42 \\ \cline{2-4} 
\multicolumn{1}{|c|}{} & Other & 8 & 1.47 \\ \hline
\multicolumn{1}{|c|}{\multirow{7}{*}{\textbf{Level of education}}} & Less than high school & 39 & 7.18 \\ \cline{2-4} 
\multicolumn{1}{|c|}{} & High school diploma or equivalent & 150 & 27.62 \\ \cline{2-4} 
\multicolumn{1}{|c|}{} & Bachelor's degree & 169 & 31.12 \\ \cline{2-4} 
\multicolumn{1}{|c|}{} & Master's degree & 134 & 24.68 \\ \cline{2-4} 
\multicolumn{1}{|c|}{} & Doctoral degree & 8 & 1.47 \\ \cline{2-4} 
\multicolumn{1}{|c|}{} & Other & 17 & 3.13 \\ \cline{2-4} 
\multicolumn{1}{|c|}{} & Did not answer & 26 & 4.79 \\ \hline
\multicolumn{1}{|c|}{\multirow{5}{*}{\textbf{Size of the household}}} & 1 & 152 & 27.99 \\ \cline{2-4} 
\multicolumn{1}{|c|}{} & 2 & 223 & 41.07 \\ \cline{2-4} 
\multicolumn{1}{|c|}{} & 3 & 80 & 14.73 \\ \cline{2-4} 
\multicolumn{1}{|c|}{} & 4 & 68 & 12.52 \\ \cline{2-4} 
\multicolumn{1}{|c|}{} & \textgreater{}4 & 20 & 3.68 \\ \hline
\multicolumn{1}{|c|}{\multirow{4}{*}{\textbf{Number of cars in the household}}} & 0 & 139 & 25.60 \\ \cline{2-4} 
\multicolumn{1}{|c|}{} & 1 & 304 & 55.99 \\ \cline{2-4} 
\multicolumn{1}{|c|}{} & 2 & 91 & 16.76 \\ \cline{2-4} 
\multicolumn{1}{|c|}{} & \textgreater{}2 & 9 & 1.66 \\ \hline
\multicolumn{1}{|l|}{\multirow{4}{*}{\textbf{Number of bicycles in the household}}} & 0 & 40 & 7.37 \\ \cline{2-4} 
\multicolumn{1}{|l|}{} & 1 & 128 & 23.57 \\ \cline{2-4} 
\multicolumn{1}{|l|}{} & 2 & 156 & 28.73 \\ \cline{2-4} 
\multicolumn{1}{|l|}{} & \textgreater{}{}2 & 219 & 40.33 \\ \hline
\multicolumn{1}{|c|}{\multirow{4}{*}{\textbf{Income}}} & Low (Up to 250.000 kr.) & 82 & 15.1 \\ \cline{2-4} 
\multicolumn{1}{|c|}{} & Medium (251-500.000 kr.) & 140 & 25.8 \\ \cline{2-4} 
\multicolumn{1}{|c|}{} & High (Over 500.000 kr.) & 221 & 40.7 \\ \cline{2-4} 
\multicolumn{1}{|c|}{} & Did not answer & 100 & 18.4 \\ \hline
\end{tabular}%
\caption{Copenhagen Sample characteristics}
\label{tab:Sample}
\end{table}
For further details on the survey and the CPH data collection and sampling processes, the reader is referred to \cite{Deliverable, Mayara}.\\
Our analysis focuses on the
data from parts 2, 4 and 5 of the questionnaire, using the latter for modelling the choice of CS subscription plans.
In this SP, each respondent answered 3 experiment tasks which presented 4 CS alternative plans with 8 associated attributes $X_n$, (one-time subscription cost, usage cost, walking time to access the vehicle, probability to get a shared vehicle, CS vehicle type, CS vehicle engine type, walking time from parking location to destination and extra features) and the option to select 'None of the alternatives.'
Figure \ref{Example} shows an example of a choice task presented to the respondents. To minimize response bias, the order of appearance of the attributes was random for each individual but was the same for the three tasks presented to the same individual.
As the name suggests, in one-way CS (station-based), or OWSB, the cars can be found in fixed locations (stations) around the city. The user can take a vehicle from one of the fixed locations available and return it to the same place or another fixed location available in the service.
In one-way CS (free-floating), or OWFF, the users can pick up or return the car in any available parking spot in the city within a delimited coverage area of the service (thus, no fixed stations), giving extra freedom to users.
In roundtrip CS (RT), one can pick up a car at fixed stations scattered around the city, and need to drop it off at the same location. Finally, peer-to-peer (P2P) CS services are those which make it possible for vehicle owners to rent out their cars to other users for a defined time period, also relying on fixed (and identical) pick-up and drop-off locations for a given ride.\\
The attributes of the alternatives included in the tasks and their corresponding levels were selected by considering the published literature and qualitative surveys previously conducted during the Share-More project \cite{Deliverable, Mayara}. The cost of usage was presented in cost per minute, hour or day to test the influence of these units on respondents' choices.
\section{Results}
\subsection{Baseline Model Results} \label{base}
\begin{figure}[H]
    \centering
    \includegraphics[width = 0.7
    \textwidth]{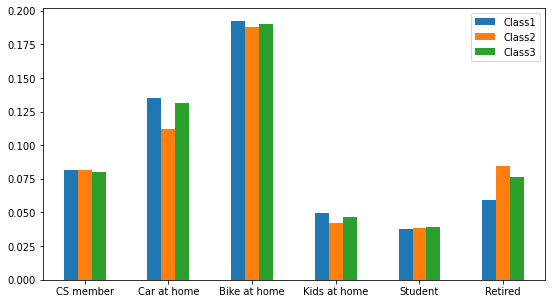}
    \caption{Representation of the class membership of the LCCM model}
    \label{Classes1}
\end{figure}

To benchmark the proposed model, we tried to use the formulation presented in the generalized random utility model \cite{walker} as a baseline. One can represent it in graphical representation, similarly as in Figure \ref{M3}, with the difference that the relation between the indicators' utility and the decision-makers socio-demographic characteristics is linear instead of a densely connected layer.
However, the class membership formulation's simplicity made the model unable to converge for our data, making the Hessian matrix for the variance computation not invertible.\\
To have a benchmark model with comparable magnitude for the likelihood, we have estimated the model with a traditional LCCM as a baseline without including the attitudinal variables. We have employed 20\% of the data for testing, dividing the data into train and test with 433 and 109 individuals, respectively. The results are presented in Table 2:
\begin{table}[H]
\centering
\resizebox{\textwidth}{!}{%
\begin{tabular}{c|l|c|c|c|c|c|c|c|c}
\textbf{Model} & \textbf{Nº Classes} & \textbf{Nº parameters} & \textbf{Null LL} & \textbf{LL} & \textbf{AIC} & \textbf{BIC} & \textbf{R-squared} & \textbf{Test null LL} & \textbf{Test LL}\\ \hline
LCCM & \multicolumn{1}{c|}{2} & 30 & -2047.21 & -1599.41 &   3258.81 & 3413 & 0.22 & -515.02 & -400.52 \\
LCCM & \multicolumn{1}{c|}{3} & 43 & -2047.21 & -1568.14 &  3234.29 & 3487 & 0.23 & -515.02 & -398.73
\end{tabular}
}
\caption{LCCM results without attitudinal variables}
\label{LCCM}
\end{table}
For the models above, we have used the same socio-characteristics for constructing the class membership as in our proposed formulation. More specifically, we employed age, binary variables that indicate if the decision-makers have a bike or car at home, their CS membership status, kids at home, or if they are students or retired.\\
However, we found that having a car at home, being retired or being a student is not statistically significant under this formulation. When applying the proposed formulation, we can effectively include this information in the model, which improve our characterization of the latent space.\\
Given the probability of each individual in the sample and its corresponding socio-characteristics, we have represented the classes in Figure \ref{Classes1}, by using the Bayes Theorem to compute:
\begin{equation}
              P_n(socio-characteristic|K=k)
\end{equation}
where $k$ is the corresponding class  number.\\
Figure \ref{Classes1} is compared in the next subsection with the proposed model results.

\subsection{Proposed Model Results}
We have explored our proposed formulation for different numbers of latent classes and latent variables. We have employed the same train/test split as for the baseline model. It is important to note that our EM process has been replicated five times with random initializations. We have computed the likelihood variance between the different obtained model estimates to check for stability. The results are summarized in the following table:
\begin{table}[H]
\centering
\resizebox{\textwidth}{!}{%
\begin{tabular}{c|c|c|c|c|c|c|c|c|c}
\textbf{Nº Classes} & \textbf{\begin{tabular}[c]{@{}c@{}}Nº latent\\ variables\end{tabular}} & \textbf{\begin{tabular}[c]{@{}c@{}}EM\\ iterations\end{tabular}} & \textbf{Null LL} & \textbf{LL} & \textbf{\begin{tabular}[c]{@{}c@{}}Variance\\ LL \end{tabular}} & \textbf{R-squared} & \textbf{\begin{tabular}[c]{@{}c@{}}Test null\\ LL\end{tabular}} & \textbf{Test LL} & \textbf{\begin{tabular}[c]{@{}c@{}}Variance\\ Test LL\end{tabular}}\\ \hline
2 & 2 & 15 & -2047.21 & -1575.32 & 14.62 & 0.23 & -515.02 &  -404.72 & 0.28\\ \hline
\textbf{3} & \textbf{2} & \textbf{30} & \textbf{-2047.21} &  \textbf{-1539.56} & \textbf{22.01}  & \textbf{0.25}  & \textbf{-515.02} & \textbf{402.72} & \textbf{9.8}\\ \hline
3 & 3 & 25& -2047.21 & -1531.41 & 26.36 &  0.25 &   -515.02 &-400.73 & 9.36 \\ \hline
\end{tabular}%
}
\caption{Model results}
\label{tab:my-table}
\end{table}
The model with three latent classes and two latent variables has given the best overall estimation. Even though the model with three classes and three latent classes has a slightly better fit, when another latent variable is included, its corresponding estimates parameters were not statistically significant, which made us reject this model formulation.\\
Comparing the results from Tables \ref{LCCM} and \ref{tab:my-table}, we observe an increase in the training likelihood for our formulation. On the other hand, we do not provide better results for the test data, but just comparable ones. This could be due to the small size of the test sample or to the fact that we don't have access to attitudinal information or $w_n$ values in the test stage, which can affect the prediction accuracy.\\
We now analyse the parameters of the best model, where all the data has been used for the estimation. \\
\begin{table}[H]
\centering
\begin{tabular}{|c|ccc|}
\hline
\multirow{2}{*}{\textbf{Variable}} & \multicolumn{3}{c|}{\textbf{Class specific choice model}} \\ \cline{2-4} 
 & \multicolumn{1}{c|}{\textbf{Class 1}} & \multicolumn{1}{c|}{\textbf{Class 2}} & \multicolumn{1}{c|}{\textbf{Class 3}} \\ \hline
$ASC_{CS\ free-floating}$ & \multicolumn{1}{c|}{3.50(0.46)} & \multicolumn{1}{c|}{-2.71(1.34)} & -1.68(0.89)\\ \hline
$ASC_{CS\ station-based}$ & \multicolumn{1}{c|}{3.04(0.48)} & \multicolumn{1}{c|}{-2.64(1.34)} & 0.07(0.80)\\ \hline
$ASC_{CS\ peer\ to\ peer}$ & \multicolumn{1}{c|}{4.12(0.52)} & \multicolumn{1}{c|}{0.47(1.45)} & 1.50(0.83)\\ \hline
$ASC_{roundtrip}$ & \multicolumn{1}{c|}{2.97(0.48)} & \multicolumn{1}{c|}{-3.50(0.40)}& 0.15(0.80)\\ \hline
$\beta_{One\ time\ subscription\ cost}$ & \multicolumn{1}{c|}{ -1.00(0.17)} & \multicolumn{1}{c|}{-1.12(0.51)}& -0.35(0.22)\\\hline
$\beta_{Usage\ cost(OWFF,OWST,RT)}$ & \multicolumn{1}{c|}{0.05(0.04)} & \multicolumn{1}{c|}{-0.16(0.09)}& -0.34(0.08)\\ \hline
$\beta_{Usage\ cost(P2P)}$ & \multicolumn{1}{c|}{-1.30(0.39)} & \multicolumn{1}{c|}{-5.57(1.31)}& -3.86(0.80) \\ \hline
$\beta_{Usage\ cost\ per\ day}$ & \multicolumn{1}{c|}{  -0.35(0.24)} & \multicolumn{1}{c|}{-2.40(0.74)} & -1.09(0.39)\\ \hline
$\beta_{Usage\ cost\ per\ hour}$ & \multicolumn{1}{c|}{  -0.10(0.21)} & \multicolumn{1}{c|}{-0.99(0.50)}& -0.43(0.34) \\ \hline
$\beta_{Only\ combustion\ cars}$& \multicolumn{1}{c|}{ -0.23(0.12)} & \multicolumn{1}{c|}{0.09(0.33)}&-0.66(0.20)\\ \hline
$\beta_{Probability\ of\ finding\ a\ shared\ car}$ & \multicolumn{1}{c|}{ 0.13(0.44)} & \multicolumn{1}{c|}{1.38(1.35)}& 2.70(0.75)\\ \hline
$\beta_{Walking\ time\ from\ parking\ to\ destination}$ & \multicolumn{1}{c|}{ -0.05(0.01)} & \multicolumn{1}{c|}{0.02(0.04)}&0.03(0.02)\\ \hline
\end{tabular}
\caption{Estimate and standard deviation of the parameters of the class-specific choice model}
\label{tab:betas3}
\end{table}
\begin{table}[H]
\centering
\begin{tabular}{|c|c|c|c|}
\hline
\textbf{Variable} & \textbf{Parameter} & \textbf{St error} & \textbf{P-value} \\ \hline
$ASC_{class_1}$ & 0.97  & 0.45 & 0.031 \\ \hline
$ASC_{class_2}$ &  -1.40 &  0.51  & 0.0057 \\ \hline
$\gamma_{kids at home, class_1}$ & 0.56 &  0.29 & 0.050\\ \hline
$\gamma_{kids at home, class_2}$ & -0.45 &  0.37 &  0.22\\ \hline
$\delta_{r_1, class_1}$ & 0.18 &  0.13  & 0.17\\ \hline
$\delta_{r_1, class_2}$ & -0.48 &  0.14 & 0.0009\\ \hline
$\delta_{r_2, class_1}$ & 0.12 &  0.10 & 0.27 \\ \hline
$\delta_{r_2, class_2}$ & -0.44 &  0.11  & 0.00\\ \hline
$b_{class_1}$ & 0.61 &  0.44 & 0.17 \\ \hline
$b_{class_2}$ & -4.014 &  0.54 & 0.00 \\ \hline
\end{tabular}%
\caption{Parameters of the class membership model}
\label{tab:class}
\end{table}
Table \ref{tab:betas3} shows the estimated parameters of the class-specific choice model with its corresponding standard deviation.
The utility for not choosing any of the CS services offered is set to zero for parameter identification.\\
Based on the values and signs of the estimated beta parameters, we observe that class 1 and class 2 are more negatively affected by the subscription cost, while class 3 is less influenced by this cost, but more negatively affected by the cost usage. 
In general, P2P presents a higher alternative-specific value than the other alternatives for all the classes. However, this is compensated by the $\beta_{Usage\ cost(P2P)}$ values being also more damaging in the three classes. Class 2 is the one less inclined about the P2P CS type since it has the lowest ASC-P2P and the more negative $\beta_{Usage\ cost(P2P)}$ estimator.
On the other hand, individuals with a high probability of belonging to class 3 are the most concerned if the type of engine is just combustion. Thus, CS could possibly be seen as an electric alternative to those who are more worried about the environmental footprint of their trips. Given the beta values for displaying the cost in hours ($\beta_{Usage\ cost\ per\ hour}$) or days ($\beta_{Usage\ cost\ per\ day}$), Table \ref{tab:betas3} also shows a bias towards displaying the price per minute (baseline) instead of per day or hour across all the classes, related to the fact that CS users tend to drive it for short time periods.
Regarding the probability of finding a car, it is a more important feature for classes 2 and 3, which make them more dependent on the availability of the service.\\
Overall, class 2 seems to be less prone to use any CS, given all its estimated parameters.

\begin{figure}[H]
    \centering
    \includegraphics[width = 0.7
    \textwidth]{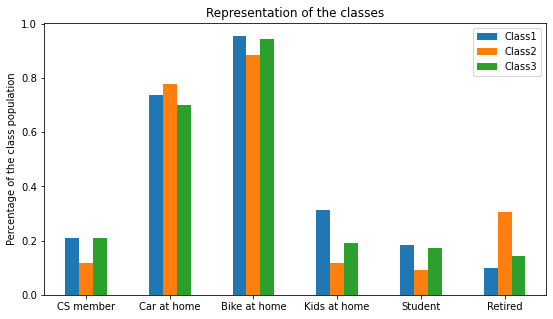}
    \caption{Representation of the class membership of our proposed model}
    \label{Classes}
\end{figure}

\begin{figure}[H]
    \centering
    \includegraphics[width = 0.7
    \textwidth]{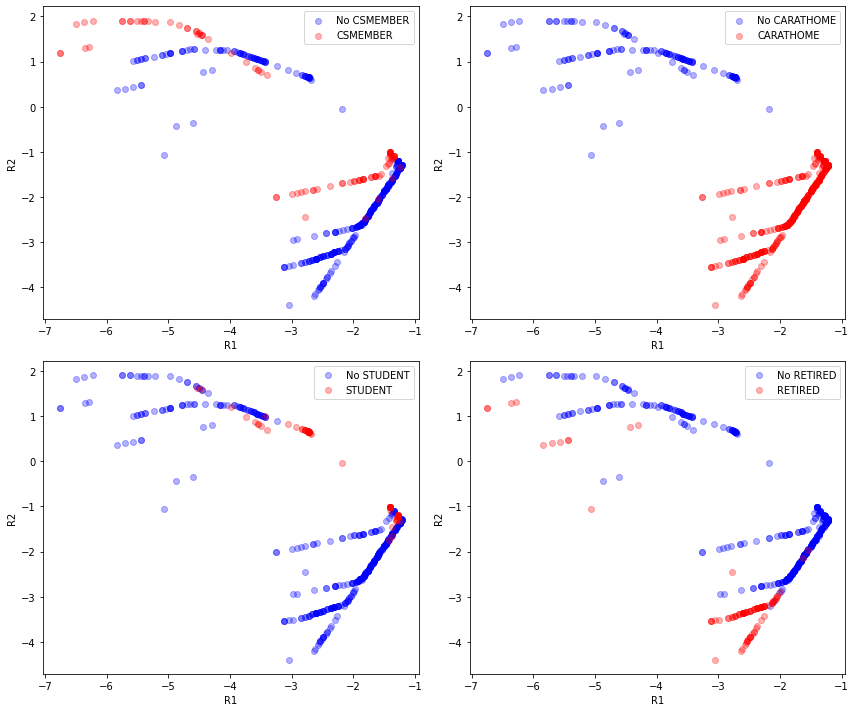}
    \caption{Latent variables representation characterised by socio-characteristics}
    \label{Latent}
\end{figure}

\begin{figure}[H]
    \centering
    \includegraphics[width = 0.6
    \textwidth]{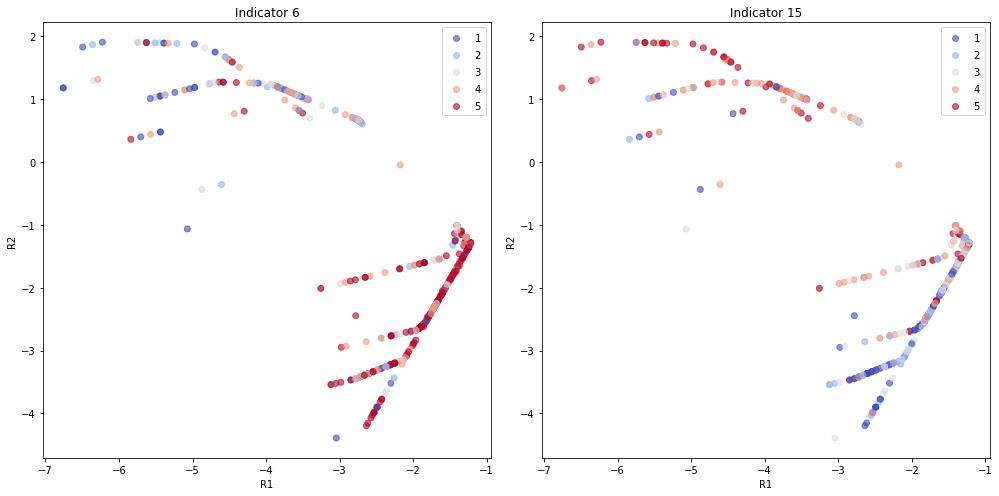}
    \caption{Latent variables representation characterized by the answers to indicators 6 and 15}
    \label{Latent1}
\end{figure}

The parameters of the class membership model are summarised in Table \ref{tab:class}. Given the probability of each individual in the sample and its corresponding socio-characteristics, we have represented the classes in Figure \ref{Classes}.
Individuals with a higher probability of belonging to class 1 have around 20\% probability of being a CS member, a bit above the sample average (17.5\%). They also tend to have more kids at home, as well as bikes than the other classes. Studies like \cite{Kids} have shown that when there are significant life changes, such as the birth of a child, people become more inclined to use CS.
On the other hand, class 2 presents the lowest probability of being a CS member and having kids or bikes at home. Retired people tend to have more predisposition for this class, while students have less. This is aligned with evidence in the literature \cite{Age} suggesting that young people are more prone to use this service.
Finally, class 3 has the same probability of being a CS member as class 1, but it is also the class with less probability of having a car at home which could make them more dependent on the availability of the service.\\
Comparison between Figure \ref{Classes1} and Figure \ref{Classes}, shows that the configuration of the classes with respect to the socio-characteristics changes when we include attitudinal information, as is expected. However, bearing in mind that the betas values of the two models are slightly different and classes cannot be directly compared.\\
Continuing with the proposed formulation results, in order to understand the implications of the coefficient of the latent variables in the utility of the class membership model, we have plotted the latent constructs for each individual in the sample and characterised them with their socio-characteristics. Figure \ref{Latent} provides us with a visual understanding of how these latent variables are constructed and distributed.
By analysing the parameters for the latent variables in Table \ref{tab:class} and looking at their distributions in Figure \ref{Latent}, we notice that the values of the first latent variable ($r_1$) are always negative. The more negative value of r1, the more probable is to belong to class 2, and therefore, the less inclined people are to use CS services. A negative value of $r_2$ seems to have the same effect. Thus, individuals with a more negative combination of $r_1$ and $r_2$ tend to be less inclined about CS and the other way around. Figure \ref{Latent}, suggests that students are more prone to use the service while retired people are the least predisposed. Moreover, having or not having a car seems to determine the clusters in which the $r_s$ values are structured, as we can see in the upper right plot of Figure \ref{Latent}.\\
The coefficients of the $\omega_n$ parameter are a representation of the heterogeneity present in the classes. Therefore, class three seems to be more heterogeneous. 
Finally, the parameters of the utility for each of the indicators are presented in Table \ref{tab:Ind}. The mean accuracy for all the indicators given the ordinal logit probability is 0.44.
\begin{table}[H]
\centering
\begin{tabular}{|c|c|c|c|}
\hline
\textbf{Indicator} & \textbf{$\alpha_1$} & \textbf{$\alpha_2$} & \textbf{$c$} \\ \hline
$I_{1}$ & 0.24& 0.42 & 0.65\\ \hline
$ I_{2}$ & 0.43 &  0.34& -0.14\\ \hline
$I_{3}$ & -0.14 & 0.28 & 0.65\\ \hline
$I_{4}$ & -0.22&  0.27&  0.87\\ \hline
$I_{5}$ & -0.40& 0.24 & 0.56 \\ \hline
$I_{6}$ & 0.19 &-0.38 & -0.62 \\ \hline
$I_{7}$ & -0.026 &  0.23 & 1.59\\ \hline
$I_{8}$ & -0.40 & 0.14 & 2.59\\ \hline
 $I_{9}$ &  0.09 & 0.045 & 1.72 \\ \hline
 $I_{10}$ & -0.070 & 0.19 & 2.56 \\ \hline
$I_{11}$ & -0.40& -0.083& 2.28  \\ \hline
$I_{12}$ & -0.33 &-0.016& 2.97 \\ \hline
$I_{13}$ & -0.040 &  0.044 & 1.80 \\ \hline
 $I_{14}$ & 0.019 & -0.10 & 2.40 \\ \hline
$I_{15}$ & -0.095 & 0.45 & 2.42  \\ \hline
 $I_{16}$ & -0.17 & 0.43 & 2.35\\ \hline
$I_{17}$ & 0.17 & 0.34 & 1.60\\ \hline
\end{tabular}%
\caption{Parameters of the measurement model}
\label{tab:Ind}
\end{table}
Finally, it is interesting to see how the answers for some indicators are distributed in the latent space. Figures \ref{Latent} and \ref{Latent1}, show that people with a car at home agree more with the statement that the car is a status symbol. For indicator 15, people with a more positive value of $r_2$ seem to agree more with the statement that they wouldn't need a car if they have CS, as we would expect given the $r_2$ coefficients of Table \ref{tab:class} and the positive $\alpha_2$ coefficient for the utility of indicator 15 in Table \ref{tab:Ind}.
\section{Conclusions}
We focused on the relevance and impact of adding attitudinal information to the models, evidencing its importance in configuring the latent classes for heterogeneity representation. We propose a new way to account for complex (and non-linear) constructs of latent classes by extending the general random utility model with ML frameworks for latent space representation.\\
We have tested our proposed framework for estimating a car-sharing (CS) service subscription choice with SP data from Copenhagen, Denmark. Our results confirm that the inclusion of attitudinal variables provides a DCM more behaviorally realistic. For example, individuals who are more inclined towards the concept of CS tend to be grouped together in classes with higher parameter estimates of the utility of choosing different CS plans. Also, as expected, such classes are sensitive to different attributes of the CS plans, but, more importantly, the complex structure of such classes can be modelled by considering non-linear formulations learned by an ANN framework. This confirms that complex beliefs and attitudes play a key role in CS subscription decision-making, and including this information allows for more accurate estimation and a better understanding of the classes.\\
From the application perspective, it provides us with extra insights to help us design better policies by having a more realistic population segmentation. In our CS application, when creating new approaches to attract and retain CS members, attitudinal information allows us to focus on changing beliefs and mindsets that we know are closely related to individual choices. Finally, it also helps to understand better how the decision-making process works from a theoretical point of view.
We have also improved our understanding of how the features of the CS business can maintain and attract new members. We have noted the importance of the subscription and usage cost throughout all the models. In addition, the price format is relevant to the decision, and the price per minute display is the most attractive, possibly driven by the past experience in the case-study at stake. We have also observed that the availability of cars is essential in attracting new members, which can be seen as an improvement opportunity for CS companies. Choosing an electric car is more favourable for those more inclined to use CS, so having a fleet with electric vehicles can be essential to maintain CS members.\\
Our study also has a few limitations. Regarding our proposed model at the estimation level, the values of the latent variables change between iterations, even though the underlying structure and model probabilities are kept between runs. This is the main difference between an ML learning approach and a traditional DCM estimation, where all the parameters need to be identifiable.
On the other hand, convergence is defined empirically by setting the number of iterations due to small fluctuations in the convergence of the EM algorithm.
Also, given the small sample size, we could not divide the dataset in training, validation, and testing; therefore, the hyperparameters referring to the number of nodes and layers in the NN were not tuned according to the validation samples.
Regarding our overall model architecture, and in order to improve the prediction performance, further work can be done, trying to apply other types of explainable AI (XAI) and other latent space representations (word embeddings) \cite{Ioanna}.\\
Although this work has its limitations, we are optimistic that this analysis has opened the door to future research on integrating attitudinal variables in DCMs through ML techniques. This investigation could improve the overall model fit and prediction accuracy, thanks to the representation of heterogeneity due to ML techniques' ability to capture complex unobserved patterns. However, we always have to bear in mind that the transparency and interpretability of new modelling frameworks are of utmost importance.
\subsection{Acknowledgements}
We thank Share-More project for providing the data used in this study and eMOTIONAL CitiesGrant agreement ID: 945307, funding from EU's Horizon 2020. \footnote{https://emotionalcities-h2020.eu/}

\newpage
\renewcommand\refname{References}


\begin{thebibliography}{99}
\bibitem{Hess} Hess, S., Ben-Akiva, M., Gopinath, D., Walker, J. (2009). Taste heterogeneity, correlation, and elasticities in latent class choice models, in Transportation Research Board 88th Annual Meeting.
\bibitem{Hess2} Hess, S. (2014). Latent class structures: Taste heterogeneity and beyond, Handbook of Choice Modelling, pp. 311–329. doi: 10.4337/9781781003152.00021.
\bibitem{walker} Walker, J., \& Ben-Akiva, M. (2002). Generalized random utility model. Mathematical social sciences, 43(3), 303-343.
\bibitem{Motoaki} Motoaki, Y., \& Daziano, R. A. (2015). A hybrid-choice latent-class model for the analysis of the effects of weather on cycling demand. Transportation Research Part A: Policy and Practice, 75, 217-230
\bibitem{SubIndicator2} Haustein, S., Hunecke, M., 2013. Identifying target groups for environmentally sustainable transport: assessment of different segmentation approaches. Curr. Opin.
Environ. Sustain.
https://doi.org/10.1016/j.cosust.2013.04.009.
\bibitem{SubIndicator3}Haustein, S., 2012. Mobility behavior of the elderly: An attitude-based segmentation approach for a heterogeneous target group. Transportation (Amst) 39,
1079–1103. https://doi.org/10.1007/s11116-011-9380-7
\bibitem{latent} Bahamonde-Birke, F.J., Kunert, U., Link, H. et al. About attitudes and perceptions: finding the proper way to consider latent variables in discrete choice models. Transportation 44, 475–493 (2017). https://doi.org/10.1007/s11116-015-9663-5
\bibitem{Carlos1} Weibo Li, Maria Kamargianni (2020) An Integrated Choice and Latent Variable Model to Explore the Influence of Attitudinal and Perceptual Factors on Shared Mobility Choices and Their Value of Time Estimation. Transportation Science 54(1):62-83.
https://doi.org/10.1287/trsc.2019.0933
\bibitem{Carlos2} Paulssen, M., Temme, D., Vij, A. et al. Values, attitudes and travel behavior: a hierarchical latent variable mixed logit model of travel mode choice. Transportation 41, 873–888 (2014). https://doi.org/10.1007/s11116-013-9504-3
\bibitem{Nobel} McFadden, D., Economic Choices. The American Economic Review, Vol. 91, No. 3 (Jun., 2001), pp. 351-378
\bibitem{Mixed} Train K.,(2009) \textit{Discrete Choice Methods with Simulation}, Cambridge University Press. 153- 168.
\bibitem{McFadden}
McFadden, D., Train, K. (2000). Mixed MNL models for discrete response. Journal of applied econometrics 15: 447-470.
\bibitem{hurtubia} Hurtubia, R., Nguyen, M. H., Glerum, A., \& Bierlaire, M. (2014). Integrating psychometric indicators in latent class choice models. Transportation Research Part A: Policy and Practice, 64, 135-146.
\bibitem{McFadden2}
McFadden, D., 1986. The choice theory approach to market research. Market. Sci. 5 (4), 275–297.
\bibitem{Indicators1} Atasoy, B., Glerum, A., \& Bierlaire, M. (2011). Mode choice attitudinal latent class: a Swiss case-study. Second International Choice Modeling Conference, July 2011
\bibitem{Likert} Likert, R. (1932) A technique for the measurement of attitudes. Arch. Psychol. 22 (140).
\bibitem{Indicators3} Krueger, R., Vij, A., Rashidi, T.H..(2016). Normative belief and modality styles: a latent class and latent variable model of travel behaviour. Springer Science+Business Media New York.
\bibitem{Indicators4} Alonso-González, M. J., Hoogendoorn-Lanser, S., van Oort, N., Cats, O., Hoogendoorn, S.
Drivers and barriers in adopting Mobility as a Service (MaaS) – A latent class cluster analysis of attitudes, (2020).
Transportation Research Part A: Policy and Practice,
Volume 132,
Pages 378-401,
ISSN 0965-8564,https://doi.org/10.1016/j.tra.2019.11.022.
\bibitem{Embeddings2} Sifringer, B., Lurkin, V., \& Alahi, A. (2020). Enhancing discrete choice models with representation learning. Transportation Research Part B: Methodological, 140, 236-261.
\bibitem{Previous1} Bentz, Y., \& Merunka, D. (2000). Neural networks and the multinomial logit for brand choice modelling: a hybrid approach. Journal of Forecasting, 19(3), 177-200.
\bibitem{Previous2} Hruschka, H., Fettes, W., Probst, M., \& Mies, C. (2002). A flexible brand choice model based on neural net methodology a comparison to the linear utility multinomial logit model and its latent class extension. OR spectrum, 24(2), 127-143.
\bibitem{Previous3} Hruschka, H., Fettes, W., \& Probst, M. (2004). An empirical comparison of the validity of a neural net based multinomial logit choice model to alternative model specifications. European Journal of Operational Research, 159(1), 166-180.
\bibitem{Ioanna} Arkoudi, I., Azevedo, C. L., \& Pereira, F. C. (2021). Combining Discrete Choice Models and Neural Networks through Embeddings: Formulation, Interpretability and Performance. arXiv preprint arXiv:2109.12042.
\bibitem{Pereira} Pereira, F. C. (2019). Rethinking travel behavior modeling representations through embeddings. arXiv preprint arXiv:1909.00154.
\bibitem{ML1} Han, Y., 2019. Neural-Embedded Discrete Choice Models.
\bibitem{Georges2} Sfeir, G., Rodrigues, F., Abou-Zeid, M. Gaussian process latent class choice models. Transportation Research Part C: Emerging Technologies, 136. https://doi.org/10.1016/j.trc.2022.103552
\bibitem{George} Sfeir, G., Abou-Zeid, M., Rodrigues, F., Pereira, F.C., Kaysi, I. (2021). Latent class choice model with a flexible class membership component: A mixture model approach. Journal of Choice Modelling, 41, 100320. https://doi.org/10.1016/j.jocm.2021.100320
\bibitem{ML2} Wong, M., Farooq, B., Bilodeau, G.A., 2018. Discriminative Conditional Restricted Boltzmann Machine for Discrete Choice and Latent Variable Modelling. J. Choice Model. 29, 152–168. https://arxiv.org/abs/1706.00505
\bibitem{MIT} Ben-Akiva, M., Walker J., Bernardino A. T., Gopinath, D. A., Morikawa, T., \& Polydoropoulou, A. (2002). Integration of choice and latent variable models. Chapter 21 In: Mahmassani, Hani S. (Ed.), In perpetual motion: Travel behaviour research opportunities and application challenges. Emerald Group
Publishing Limited, Bingley, United Kingdom, pp. 431–470.
\bibitem{Train} Train, K.E., 2008. EM algorithms for nonparametric estimation of mixing distributions. J. Choice Model. 1, 40–69. https://doi.org/10.1016/S1755-5345(13)70022-8
\bibitem{EM} Dempster, A.P., Laird, N.M., Rubin, D.B., 1977. Maximum Likelihood from Incomplete Data via the EM Algorithm. J. R. Stat. Soc. Ser. B 39, 1–38. https://doi.org/10.1177/019262339101900314
\bibitem{Sequencial2} Train, K., D. McFadden and A. Goett (1986). The Incorporation of Attitudes in Econometric Models of Consumer Choice. Cambridge Systematics working paper.
\bibitem{Nocedal} Nocedal, J., Wright, S.J., Robinson, S.M., 1999. Numerical Optimization. Springer, New York.
\bibitem{Deliverable} Frenkel, A., Shiftan, Y., Gal-Tzur, A., Tavory, S. S., Lerner, O., Antoniou, C., Cantelmo, G., Amini, R. E., Lima Azevedo, C. M., Monteiro, M.M., Kamargianni, M., Shachar, F. S., Israel, D., Behrisch, C., Schiff, K., Shalev, J., \& Peretz, D.(2021). Share More: Shared MObility Rewards - Summary report. https://orbit.dtu.dk/en/publications/share-more-shared-mobility-rewards-summary-report
\bibitem{Mayara} Monteiro, M. M.; Azevedo, C. M. L.; Kamargianni, M.; Cantelmo, G.; Tavory, S. S.; Gal-Tzur, A.; Antoniou, C.; Shiftan, Y. Car-Sharing Subscription Preferences and the Role of Incentives: The Case of Copenhagen, Munich, and Tel Aviv-Yafo. arXiv preprint arXiv:2206.02448
\bibitem{bike} Københavns Kommune. City of Cyclists. https://urbandevelopmentcph.kk.dk/artikel/city-cyclists Accessed Apr. 30, 2020.
\bibitem{Kids} Priya Uteng, T., T. E. Julsrud, and C. George. The Role of Life Events and Context in Type of Car Share Uptake: Comparing Users of Peer-to-Peer and Cooperative Programs in Oslo, Norway.
Transportation Research Part D: Transport and Environment, Vol. 71, No. June 2018, 2019, pp.186–206. https://doi.org/10.1016/j.trd.2019.01.009.
\bibitem{Age} Prieto, M., G. Baltas, and V. Stan. Car Sharing Adoption Intention in Urban Areas: What Are the Key Sociodemographic Drivers? Transportation Research Part A: Policy and Practice, Vol. 101, 2017, pp. 218–227. https://doi.org/10.1016/j.tra.2017.05.012.
\end{thebibliography}
\end{document}